\documentclass[10pt, conference]{IEEEtran}

\usepackage{multirow}
\usepackage[table]{xcolor}
\usepackage{graphicx}
\usepackage{cite}
\usepackage[hidelinks,bookmarks=false]{hyperref}
\usepackage{balance}

\ifCLASSINFOpdf
\else
\fi

\begin{document}

\title{Ticket Coverage: Putting Test Coverage into Context}


\author{\IEEEauthorblockN{Jakob Rott}
\IEEEauthorblockA{Technische Universit\"at M\"unchen\\
Munich, Germany\\
rott@fs.tum.de}
\and
\IEEEauthorblockN{Rainer Niedermayr}
\IEEEauthorblockA{University of Stuttgart, CQSE GmbH\\
Stuttgart, Germany\\
niedermayr@cqse.eu}
\and
\IEEEauthorblockN{Elmar Juergens, Dennis Pagano}
\IEEEauthorblockA{CQSE GmbH\\
Garching b. M\"unchen, Germany\\
\{juergens,pagano\}@cqse.eu}
}

\maketitle

\begin{abstract}
There is no metric that determines how well the implementation of a ticket has been tested.
As a consequence, code changed within the context of a ticket might unintentionally remain untested and get into production.
This is a major problem, because changed code is more fault-prone than unchanged code.
In this paper, we introduce the metric \textit{ticket coverage} which puts test coverage into the context of tickets.
For each ticket, it determines the ratio of changed methods covered by automated or manual tests.
We conducted an empirical study on an industrial system consisting of 650k lines of Java code
 and show that ticket coverage brings transparency into the test state of tickets and reveals relevant test gaps.
\end{abstract}


\IEEEpeerreviewmaketitle

\section{Introduction}
\label{Sec_Introduction}


In long-lived systems, bugs typically occur in code areas that have been recently changed \cite{2005-use-of-relative-code-churn-measures} \cite{2005-top-ten-list}.
As a consequence, test managers put great emphasis on thoroughly testing modified code.


However, with modern agile development processes that endorse short release cycles, for large systems it is nearly impossible to execute the complete suite of tests before a release---especially if it contains a large number of manual tests. 

As changes made since the last test phase are presumably more bug-prone than unchanged parts, selecting test cases that execute the changed code would help to narrow down the scope and save resources.
But without suitable analyses it is difficult to reliably identify changed code
 and to determine which changed code chunks have been covered by automated or manual tests at a given point in time.

To make things even more complicated, tests often have to be performed with development still ongoing during an iteration.
Since the documentation of requirements is often scarce,
 testers often do not know exactly what they need to test and discount special cases by mistake. 

It is therefore not surprising that a substantial amount of changes reaches production untested, despite systematically planned and executed testing processes \cite{2013-did-we-test-our-changes}.


So what can be done to systematically focus testing efforts on changed code in agile development projects?

Since development activities are typically guided by user stories, a meaningful way to narrow down the scope is to focus testing on the changes made within the course of a given user story.
Usually, these changes are documented in \emph{tickets} that eventually correspond to user stories---either directly or by means of aggregation.

Tickets are written in natural language and therefore also understandable to non-coders, such as testers.
But since a ticket does not necessarily describe all special cases that the developer implemented, a tester typically will not be able to cover all changes given the ticket alone.

Fortunately, in many systems there is a reliable connection between tickets and code changes via commit messages, such that code areas changed due to a ticket can be identified. 


Our contribution is a new metric, \emph{ticket coverage}, that unveils, which of the changes made in the course of a given ticket are left untested.
To investigate the usefulness of the metric in practice, we conducted a first empirical study.
We show that ticket coverage yields meaningful results and that it enables testers to reveal important test gaps.
Moreover, we present how the metric can be further improved by systematically excluding unimportant test gaps with certain characteristics from the computation.


The remainder of this work is organized as follows.
In Section \ref{Sec_Terms} we define important terms and introduce ticket coverage.
Section \ref{Sec_Related_Work} discusses related work.
Section \ref{Sec_Approach} describes our approach to measure ticket coverage.
In Section \ref{Sec_Study} we report on the empirical study we performed to investigate its usefulness.
Finally, Section \ref{Sec_Conclusion} concludes the paper with our ideas for future work.

\section{Terms and Definitions}
\label{Sec_Terms}
This section explains terms used in this paper and introduces the \textit{ticket coverage} metric.

\medskip

A \textit{ticket} documents a requested feature or a reported bug.
It is written in natural language and stored in a ticket tracking system.
The proposed metric requires a ticket-based implementation process.

\medskip

We call methods that were added or changed during the implementation of a ticket and not tested afterwards \textit{test gaps}.
Eder et al. showed that bugs occur more likely in changed-untested methods than in others \cite{2013-did-we-test-our-changes}.

\medskip

Test coverage data is often not limited to coverage that stems from the execution of test cases,
 but also includes coverage recorded during \textit{startup routines},
 which are executed unconditionally at each startup of a program.
We use the term \textit{test-dependent coverage} to refer to coverage recorded during the (actual) execution of test cases,
 and \textit{test-independent coverage} to refer to coverage recorded during startup routines prior the test execution.
We distinguish between these two coverage sources and report the values separately, because we consider test-independent coverage to be less meaningful.

\medskip

\begin{figure}
	\centering
	\includegraphics[width=6cm]{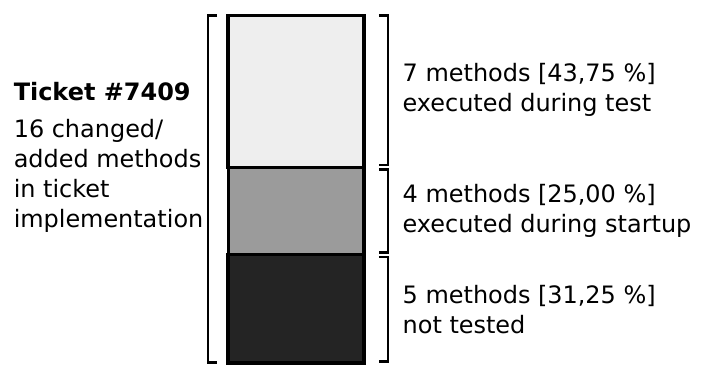}
	\caption{Representation of ticket coverage for a sample ticket as stacked bar chart. 16 methods were added or changed during the implementation of the ticket. 11 methods got executed during the test execution; 7 exclusively through the test case (light gray), 4 of them test-independently during the program startup (dark gray). 5 methods were not tested (black).}
	\label{Fig_TicketCoverageTrafficLight}
\end{figure}

The metric \textit{ticket coverage} gives information about the test state of a ticket.
It relates to the methods that were added and changed during the implementation of a given ticket,
 and expresses which ratio of these methods was tested.
The proposed metric uses coverage data at the granularity level of methods\footnote{The
 coverage at the method level considers a method as covered if at least one of its statements is executed.},
 as methods are the smallest named executable entities.
Ticket coverage is easily interpretable without code knowledge by visualizing the amount of covered methods that were changed during the implementation a ticket.
Tables or stacked bar charts can be used to present the results (see example in Figure \ref{Fig_TicketCoverageTrafficLight})
 and enable test managers to quickly get an overview over the testing state of important tickets.

\section{Related Work}
\label{Sec_Related_Work}

The idea to compute \textit{ticket coverage} and use it as a metric is new to the best of our knowledge.

Related work exists in the area of code coverage, selective regression testing, and defect prediction.

\medskip

\textbf{Code coverage} metrics and test adequacy criteria have been a major research focus for the last decades \cite{zhu1997software}. 
Code coverage metrics express which ratio of the code of a software is executed when running the test cases.
Much effort has been spent in assessing the relationship between code coverage and test suite effectiveness \cite{inozemtseva2014coverage} \cite{andrews2006using} \cite{namin2009influence} \cite{niedermayr2016teststellme}.
The metrics can be computed at different levels (e.g., at the method, line, or branch level) and are widely used in practice.
To our knowledge, code coverage has not yet been linked to tickets and computed for the changes conducted in the implementation of a ticket.

\medskip

\textbf{Selective regression testing} makes use of techniques to derive an appropriate subset of existing test cases with the goal to reduce the test effort.
Most techniques perform their selection based on information about the code, modifications, and code coverage
 \cite{yoo2012regression} \cite{rothermel1996analyzing}.
However, selective regression testing does not work well in practice.
Graves \cite{graves2001empirical} showed that minimalized test suites created with regression test selection techniques
 yield only equivalent fault detection results to slightly larger test suites created by randomly selecting test cases.
Another study conducted at Wincor Nixdorf \cite{regression2011wincor} confirmed that this technique is not suitable to significantly reduce the effort for manual tests.

While selective regression testing tries to reduce an existing test suite such that, e.g., all changed code (and code invoking changed code) is still covered,
we identify untested changes made in the course of a specific ticket.
Therefore, our work focuses on assessing the completeness of the covered code after the test execution,
 while selective regression testing is applied before the test execution to determine the relevant test cases.
Moreover, our approach deals with changed methods but does not consider the control flow or dataflow.

\medskip

\textbf{Defect prediction} uses code characteristics, change metrics, and information about past defects to build models that can predict fault-prone code areas.
Most prediction models are built for predictions at the component or class level
 \cite{nam2014survey} \cite{kamei2016defect},
recent studies also proposed predictions at the method level
 \cite{giger2012method} \cite{hata2012bug}.
Eder et al. suggested that the probability of bugs is increased in changed-untested methods \cite{2013-did-we-test-our-changes}.
We do not predict faults or fault-prone areas, but we focus on code modifications which have an increased probability of bugs
 if they are untested (compared to unchanged code).
We aggregate the untested changes at the level of tickets.

Sherlund \cite{sherlund1995modification} developed a prototype which focuses on testing program modifications
 and requires that modified code must be executed in order to satisfy the test criteria.
While the set-up is comparable to the one used in our work, the major difference lies in the scope of the changed code.
Sherlund computed changes by comparing the most recent code version with the code at a certain reference point,
 while we compute the changes for a specific ticket by investigating all associated commits.
\section{Approach}
\label{Sec_Approach}

\begin{figure}
	\centering
	\includegraphics[width=6cm]{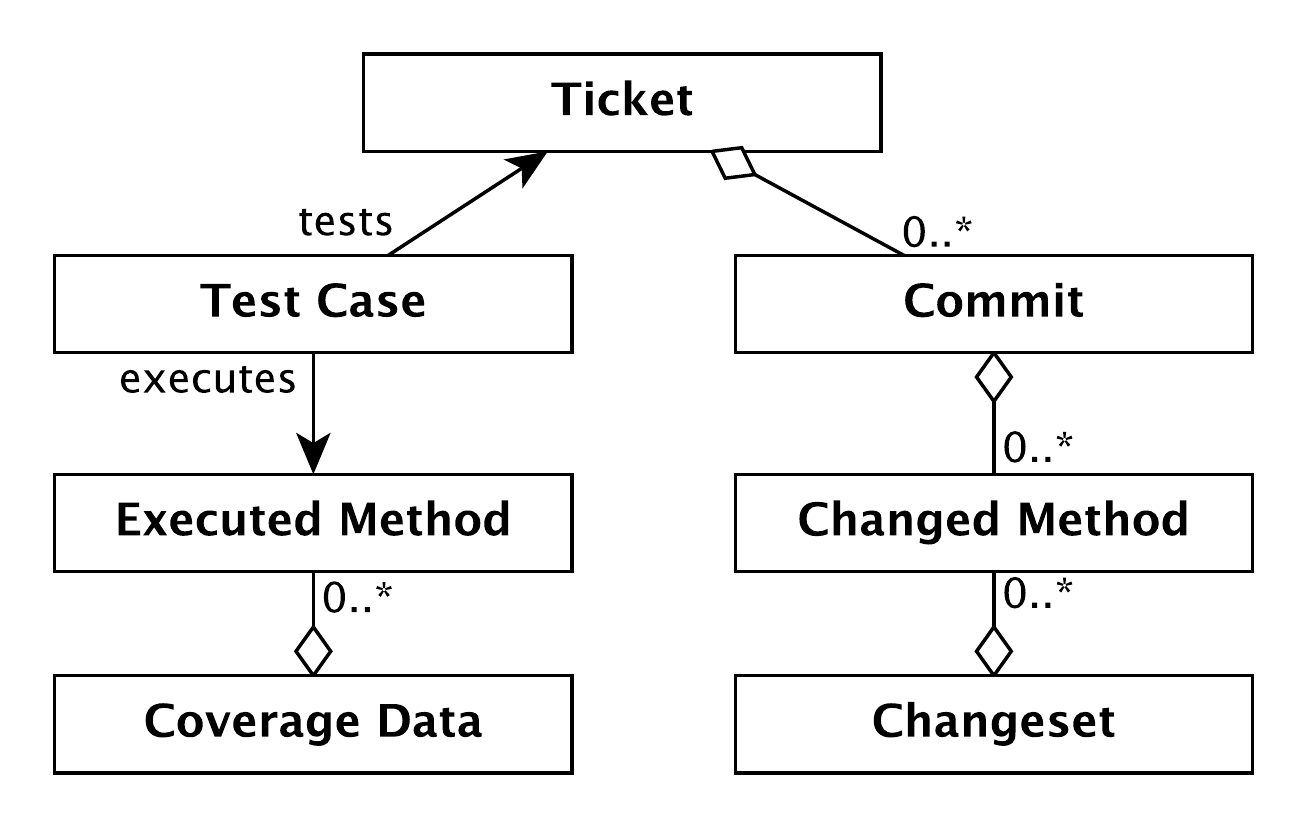}
	\caption{Relation between ticket, method changeset and coverage.}
	\label{Fig_TicketCoverageCalcModel}
\end{figure}

Ticket coverage integrates the set of added or changed methods during the implementation of a ticket
 with method-based coverage information recorded during the test execution.
Figure \ref{Fig_TicketCoverageCalcModel} shows the relation between ticket, method changeset, and coverage.

\subsection{Identify the method changeset of a ticket}
When working on a ticket, developers commit changes to the source code into the version control system.
It is common practice to include the ticket number in the commit message.

To determine the method \emph{changeset} of a ticket,
 we first fetch all commits from the version control system whose message contains the number of the particular ticket.
Then, we retrieve the changed files from each of these commits and parse the files using a shallow parser.
That allows us to identify the methods in the files.
Next, we use the commit diff to filter the methods so that only methods with changed lines remain.
Finally, we accumulate the obtained methods of all commits associated with the ticket.
The resulting set includes methods that got added or changed during the implementation.
Methods that existed prior to the ticket implementation and got deleted or existed only temporarily (added and deleted within the ticket implementation) may also exist, but they are not relevant for the computation of the ticket coverage.\footnote{Note
 that the removal of a method that prior overrode a method in a super class can influence the logic.
 However, ticket coverage is based on methods and not on the control flow.}

\subsection{Gather coverage data}
\emph{Coverage data}, the second component of ticket coverage, is obtained by executing automated or manual tests.
Profilers are attached to the program execution and record which methods get executed.

Ticket coverage is more meaningful if the coverage data is obtained from the actual test execution.
If it is not possible to record test-exclusive coverage in isolation,
 test-independent coverage data from the execution of startup routines might be included.
Test-independent coverage does not allow us to infer anything for a specific test case and needs to be treated differently.
To identify methods with test-independent coverage, we collect coverage once for the application startup.
\section{Study}
\label{Sec_Study}

This section reports on the empirical study that we conducted on a real-world software project
 to investigate the usefulness of ticket coverage.

\subsection{Study Object}
\label{Sec_Study_Object}

We selected \textit{Teamscale}\footnote{\url{https://www.teamscale.com}} as study object.
Teamscale is a software solution for the continuous quality analysis of program source code.
It is designed as a client-server application:
 the front-end is written in JavaScript and consists of about 90k lines of code,
 the backend is implemented in Java and comprises about 650k lines of code.

At the time when the study was conducted, a SVN repository was used to manage the source code.
The repository contains the source-code history of more than 11 years in more than 70,000 commits.
All tickets are managed in Redmine\footnote{\url{http://www.redmine.org}}
 and each ticket is classified either as \textit{Feature}, or \textit{Bug}, or \textit{Maintenance}.
Each commit message starts with the ticket number (this is enforced by an Eclipse plugin),
 therefore it is possible to draw a connection between a commit and the corresponding ticket.

The development team has about 30 developers on average and
 consists of both very experienced developers and students.
The team follows a strict development process \cite{cqse-dev-process}
 which specifies coding guidelines, stipulates automated tests and involves a reviewing process.
It is worth mentioning that developers are encouraged to improve nearby code that contains minor structural flaws
 when performing changes in the same artifact.
As a consequence, code of methods that are not directly related to the processed ticket may get changed
 and appear as changed within that ticket.
\subsection{Research Questions}
In this study, we examine the \textit{ticket coverage} metric.
We want to find out whether this new metric is suited to reveal relevant test gaps
 (i.e., added and changed untested methods in the context of tickets)
 and identify its strengths and weaknesses.
We study the following three research questions:

\medskip

\textbf{RQ1: How many of the detected test gaps are relevant for developers?}
We assume that some of the detected test gaps are relevant, represent a risk, and should be closed by extending test cases,
 while other gaps may be less relevant and not necessarily worth testing.
Therefore, we want to find out how many test gaps are relevant from the view point of developers
 to understand the benefit of this metric.

\medskip

\textbf{RQ2: Why are some test gaps irrelevant and can they be excluded systematically?}
We investigate characteristics of test gaps that are considered as irrelevant
 to identify indicators for automatically detecting these test gaps.
The answer to this research question helps improve the precision of the metric by filtering out irrelevant test gaps.

\medskip

\textbf{RQ3: How much ticket coverage is independent of the actual test case?}
We want to find out the ratio of the measured coverage that originates from the actual test cases
 as well as the ratio that is test-independent because it results from the startup of the system.
We consider coverage gained during the startup as much less expressive 
 because testers do not (systematically) perform any ticket-specific checks (actual vs. expected behavior) during this phase.
Therefore, we need to know the impact of test-independent coverage to understand the validity of the ticket coverage values.

\subsection{Study Design}

For the study, we randomly picked 54 tickets based on these criteria:
\begin{itemize}
  \item the implementation of the ticket was conducted within the last 20 months to increase the likelihood that developers remember their work
	\item code changes of the ticket involved Java code (i.e., tickets that affected only JavaScript code were excluded\footnote{Our
	  current implementation does not support coverage data from JavaScript profilers. We reserve this for future work.})
	\item at most 3 tickets from the same developer were picked to avoid developer or task specific effects
\end{itemize}
 
The selected tickets consisted of 37 feature tickets, 10 bug tickets, and 7 maintenance tickets.
9.46 Java methods were added or modified on average in the selected tickets.

For each of the 54 chosen tickets, we wrote a specification for a manual test case 
 and asked the assignee of the ticket to validate the specification to ensure that it was sufficient and proper.
Then, we executed the test case on a version of the system under test that was built from the code base at the ticket completion timestamp.
While executing the test case, JaCoCo\footnote{\url{http://www.eclemma.org/jacoco}}, which was attached to the JVM as a Java agent, recorded the coverage data at the method level.
If a test case required the study object to be in a specific state (e.g., architecture-analysis completed), we prepared this state in advance without recording coverage for that.
Since the JaCoCo coverage data included the whole run of the JVM for each built version,
 the startup coverage was determined by recording the coverage of the system initialization itself.

Then, we computed the ticket coverage for each ticket as presented in \ref{Sec_Approach}.

\medskip

The study design of the research questions is as follows:

\medskip

\textbf{RQ1:} We discussed the test gaps that were found for the investigated tickets with the corresponding assignees.
We conducted a survey in which the developers were asked to rate the findings either as relevant or irrelevant.
Subsequently, we analyzed the ratio of test gaps that were rated as relevant.
Then, we categorized the justifications of developers why they considered certain test gap as not relevant.

\medskip

\textbf{RQ2:}
We collected all methods that were added or changed but not executed during the test case.
Then, we divided these test gaps into important findings and potentially unimportant findings.
For the unimportant findings, we created categories with distinguishing characteristics and provided a suggestion for each one on how to detect and exclude its methods automatically using static analysis.

The classification in important and potentially unimportant was verified in RQ1.

\medskip

\textbf{RQ3:}
To answer this question, we calculated the ratio of covered methods in the startup coverage to all covered methods (of the ticket) to find out how much ticket coverage is independent from the actual test.

\subsection{Results}

In total, 511 methods (avg. per ticket: 9.46) got added or changed during the implementation of the 54 analyzed tickets.
Out of these, 110 methods (avg. per ticket: 2.04) were not executed during the coverage recording and therefore represent test gaps.
Out of the remaining 401 covered methods, 364 were exclusively executed by the test case and 37 methods were already executed during program initialization.
The results of the research questions are:

\medskip

\textbf{RQ1: How many of the detected test gaps are relevant for testers?}
Table \ref{Tab_TestGapCategoriesGivenByDevs} presents the results. 
20 methods (18.2\%) out of the 110 untested methods found were rated as critical by the developers. 
56 (50.9\%) were rated as uncritical, 3 (2.7\%) were rated as not coverable by manual tests,
 and 28 (25.5\%) were rated as not interesting.
RQ2 provides more details on the developers' reasons for this statement.

In two cases, the developers answered that the untested method should have been executed during the test case.
A check revealed that those 2 methods were no false positives, but the developer was wrong:
In the first case, we discovered that the test case, which had previously been verified as complete, was missing a specific ticket-relevant case.
In the second case, we found a comment indicating that the untested method was pre-implemented for later usage and not in use at that time.

Even though all test cases were verified as proper and sufficient for the
respective tickets as mentioned above, the developers answered in 18
cases that the revealed test gap method was not executed because of the test case definition. 
For example, once, an analysis execution should have been part of the test case and not a precondition.
According to the test case definition, we inspected only the results of the analysis (and recorded the coverage information for that),
 but this was not sufficient to cover the implementation changes.

The largest category of untested methods is built up of 54 methods that were
rated as not ticket relevant (i.e., undertaken changes are not related to the ticket) methods or as refactorings.
The developers gave clarifying comments for 31 of these methods why they are not ticket relevant or not supposed to be executed as they were not changed semantically.
We categorized the reasons and assigned each method to one or more of the following categories:

\begin{itemize}
\item 15 methods: removal of throws declaration led to removal of throws declarations of all caller methods
\item 8 methods: method was renamed
\item 7 methods: value that got used multiple times in a method was extracted to a constant
\item 5 methods: method got extracted from an existing method and reused in multiple locations
\item 2 methods: method parameter or a return type changed and caller methods had to be adapted
\end{itemize}

In one quarter of the analyzed tickets with untested methods, the developers rated at least two-thirds
of the test gaps as worth testing. In total, 39 (35.5\%) of the 110 test gaps found were rated as worth testing
within the scope of the analyzed ticket.

\begin{table}
 \centering
 \caption{Categorization of test gaps by developers.}
 \cellcolor{gray!25}
 \scriptsize
 \renewcommand{\arraystretch}{1.5}
  \begin{tabular}{p{2.3cm}p{3.6cm}rr}
	  \multicolumn{1}{c}{\textbf{Category}} & \textbf{Sub Category} & \multicolumn{2}{c}{\textbf{Frequency}} \\
	  \hline
\multirow{2}[2]{*}{\parbox{2.3cm}{\centering\bfseries critical:\\20/110 (18,2\%)}} &  \cellcolor{gray!25}should be executed &  \cellcolor{gray!25}2 &  \cellcolor{gray!25}1,8\%\\
	\cline{2-4}
	& uncomplete test case & 18 & 16,4\%\\
	\hline

	\multirow{3}[3]{*}{\parbox{2.3cm}{\centering\bfseries rather uncritical:\\ 56/110 (50,9\%)}} &  \cellcolor{gray!25}not ticket relevant or refactoring &  \cellcolor{gray!25}54 &  \cellcolor{gray!25}49,1\%\\
	\cline{2-4}
	& exception thrown by fatal errors & 1 & 0,9\%\\
	\cline{2-4}
	&  \cellcolor{gray!25}overridden method & \cellcolor{gray!25} 1 & \cellcolor{gray!25} 0,9\%\\
	\hline

\multirow{2}[2]{*}{\parbox{2.3cm}{\centering\bfseries need other co\-verage:\\ 3/110 (2,7\%)}} & IDE integration code & 1 & 0,9\%\\
	\cline{2-4}
	&  \cellcolor{gray!25}method for unit test & \cellcolor{gray!25} 2 &  \cellcolor{gray!25}1,8\%\\
	\hline

\multirow{3}[3]{*}{\parbox{2.3cm}{\centering\bfseries uninteresting according to RQ2: \\28/110 (25,5\%)}} & simple getter& 12 & 10,9\%\\
	\cline{2-4}
	&  \cellcolor{gray!25}too trivial to test &  \cellcolor{gray!25}12 &  \cellcolor{gray!25}10,9\%\\
	\cline{2-4}
	& toString method & 4 & 3,6\%\\
	\hline
	
	&  \cellcolor{gray!25}answer missing &  \cellcolor{gray!25}3 &  \cellcolor{gray!25}2,7\%\\
	\hline
	
	& $\Sigma$ & 110 & 100\%
  \end{tabular}
\label{Tab_TestGapCategoriesGivenByDevs}
\end{table}

\medskip

\textbf{RQ2: Why are some test gaps irrelevant and can they be excluded systematically?}
Out of the 110 test gaps found, we considered 28 methods as not worth testing due to their characteristics.
Table \ref{Tab_IrrelevantTestGaps} shows the categories of these methods.

We gathered implementations of the \texttt{\footnotesize{toString}} method in one category as we consider those as uncritical test gaps. Since every implementation overrides the \texttt{\footnotesize{toString}} method of the \texttt{\footnotesize{Object}} class, all of them can easily be found automatically.

13 methods were simple getters which return a value and do not contain any additional logic.
A simple getter can easily be detected automatically, because it is named after the member to be returned, does not to have any parameters and consists of only a single statement.

We considered methods that are ``too trivial to test'' as a further category for irrelevant test gaps.
For the classification, the number of statements was decisive.
The category contained 7 constructors that invoked only the super class constructor with at most one additional variable assignment (method length: 1 to 2 statements),
 and 5 methods that were returning only a boolean value (method length: 1 statement).

\begin{table}
\centering
\caption{Methods that were considered as irrelevant for testers were separated in three categories.}
 \scriptsize
 \renewcommand{\arraystretch}{1.5}
	\rowcolors{2}{gray!25}{white}
	\begin{tabular}{lr}
	\textbf{Category} & \textbf{Frequency} \\
	\hline
	toString method & 4 (3.6\%) \\
	simple getter & 12 (10.9\%) \\
	``too trivial to test'' & 12 (10.9\%) \\
	\hline
	$\Sigma$ unimportant methods & 28 (25.4\%)
\end{tabular}
\label{Tab_IrrelevantTestGaps}
\end{table}

With this research question, we wanted to find out in a manual analysis which categories of irrelevant test gaps appear and how many methods are assigned to them.
The presented methods can be detected and excluded automatically in order to improve the ticket coverage computation. 

\medskip

\begin{figure}[b]
	\centering
	\includegraphics[width=8.8cm]{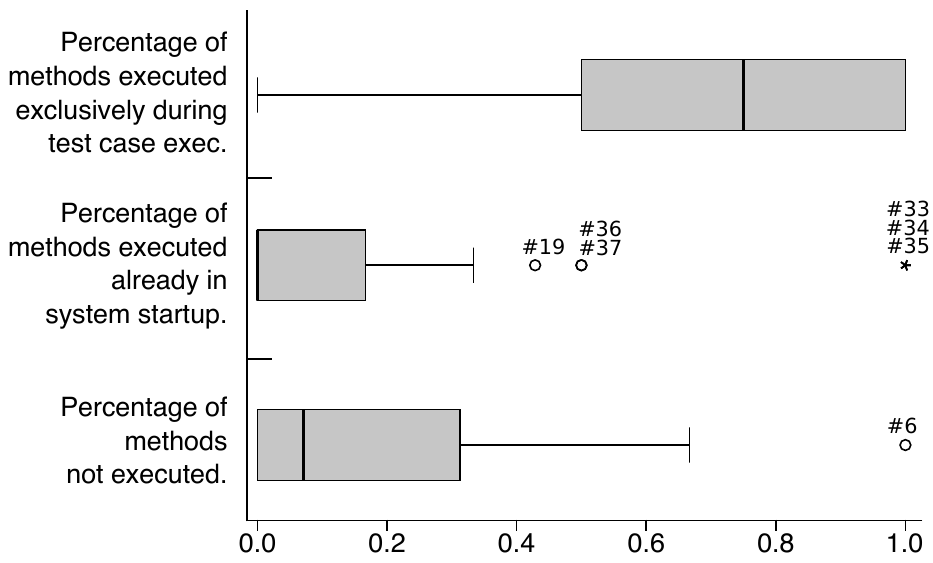}
	\caption{Boxplots that visualize the percentages of coverage during startup and test 
		and the percentages of changeset methods not executed.}
	\label{Fig_BoxplotPercentagesMethodExecution}
\end{figure}

\textbf{RQ3: How much ticket coverage is independent of the actual test case?}
The boxplots in Figure \ref{Fig_BoxplotPercentagesMethodExecution} give an overview over the execution distribution of the method changesets.
Out of the 401 changed and tested methods, only 37 (9.2\%) had been executed during the startup of the system.

\begin{figure}
	\centering
	\includegraphics[width=8.8cm]{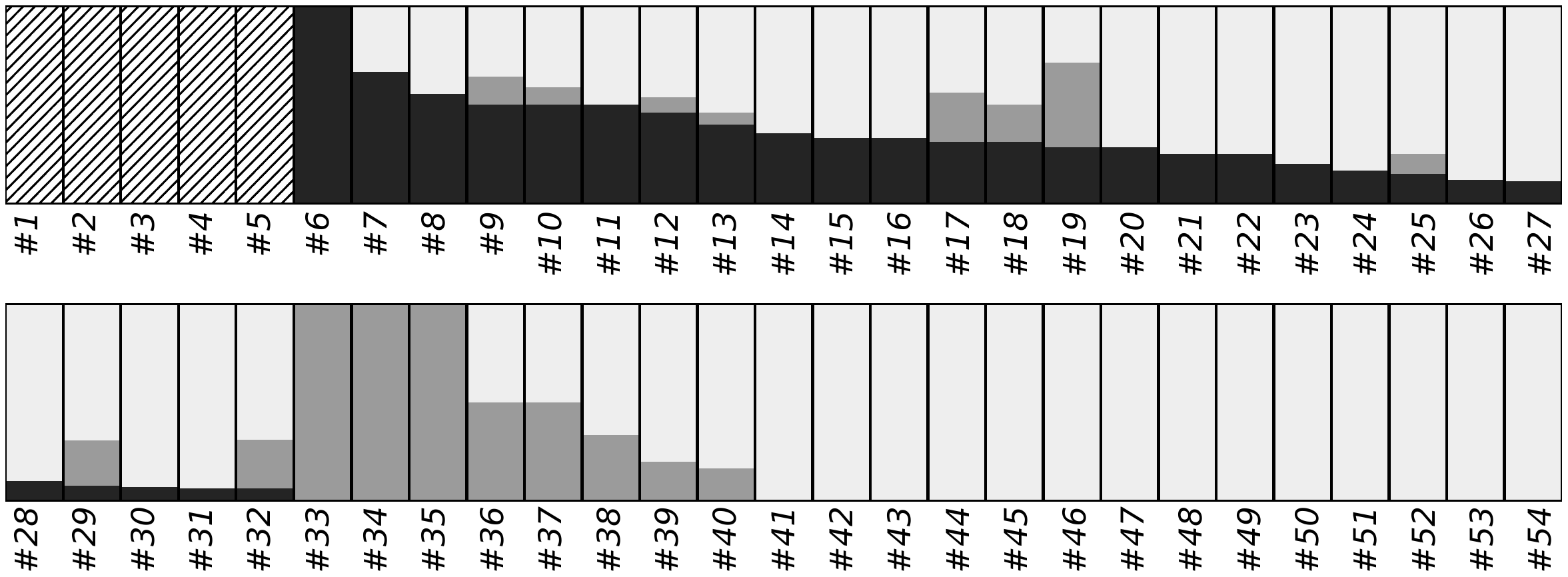}
	\caption{Overview over the ratio of methods either covered exclusively through the test case (light gray) or the system's startup (dark gray). Fraction of not executed methods is filled in black. The shaded boxes (\#1-\#5) show tickets involving only changes to unit test code, which was not executed in this study.}
	\label{Fig_TrafficLightsAllTickets}
\end{figure}

Figure \ref{Fig_TrafficLightsAllTickets} shows the ratio of methods covered by the test cases and the ratio of methods executed during the system startup.
The coverage of half of the tickets did not contain any startup coverage.
In 75\% of the executed methods, the ratio of test-independent coverage does not exceed 18\%.
The average ratio of startup coverage is 13.3\% per ticket. The average ratio of startup coverage in the tested parts of the tickets is 15.2\%.

We discuss two outliers, which were completely covered during startup routines or not covered at all, in more detail.

Ticket \#33 was fully covered during the startup. In this ticket, a huge JavaScript implementation
took place and only a few Java lines were changed to register the new JavaScript files.
This changed method is executed at each start of the system.

Ticket \#6 remained completely uncovered. In this ticket, a functionality was implemented that handles
special kinds of connection errors.
When performing the tests, we were not able to reproduce the errors in order to execute the error handling code.

The first 5 tickets displayed in Figure \ref{Fig_TrafficLightsAllTickets} involved only changes to methods
  which were exclusively reachable by unit tests.
As we did not execute unit tests, no coverage information existed for those tickets.

\subsection{Interpretation}
\textbf{RQ1:}
According to the developers, ticket coverage allows us to find relevant untested methods. 
18.2\% of the untested methods were assigned to the critical category (see Table \ref{Tab_TestGapCategoriesGivenByDevs}),
 because the developer of the ticket either thought the method should have been covered or the developer recognized that the test case was incomplete.

The high number of methods that contained ticket-irrelevant changes or were refactorings could be traced back to the performed development process that encourages nearby code improvements
 (refer to Section \ref{Sec_Study_Object} for details).
Therefore, the metric computation could be improved by detecting and filtering methods that were changed only due to refactorings (which do not change the semantical behavior of the code).

As we focused only on manual tests performed on the web interface, we did not consider coverage that could be gained from unit tests or by testing the IDE plugin.
Integrating additional coverage would be feasible.

\medskip

\textbf{RQ2:}
To improve the usefulness of the ticket coverage as metric, it is important to exclude irrelevant test gaps.
What we did manually in this study can be automated for the proposed categories.

Simple getter methods can be identified using static analysis.
For methods that were considered as too trivial to test, a simple count of the statements could be used to detect these.
\texttt{\footnotesize{toString}} methods override the \texttt{\footnotesize{toString}} implementation of the \texttt{\footnotesize{Object}} class and can therefore easily be identified.

\begin{figure}
	\centering
	\includegraphics[width=8.8cm]{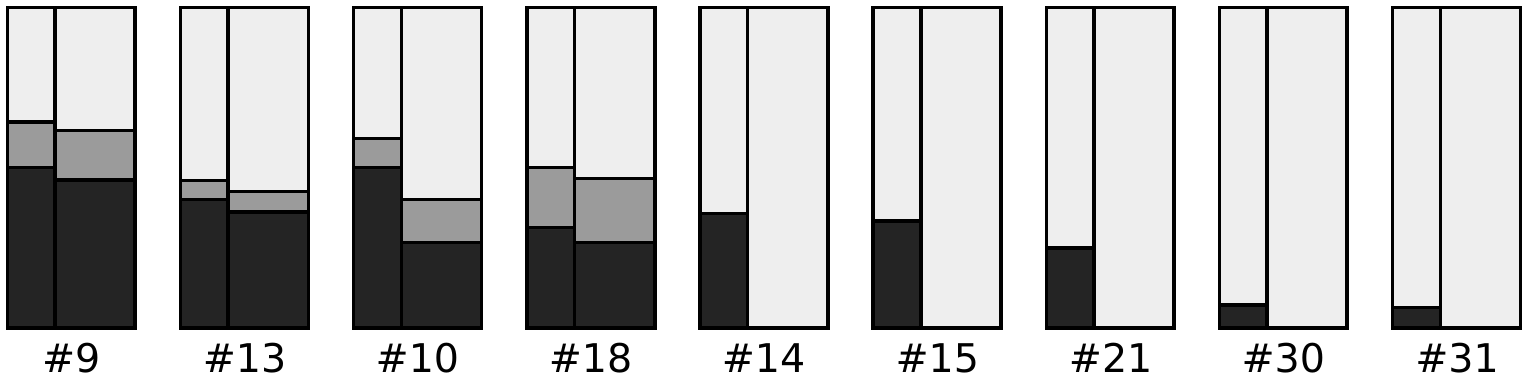}
	\caption{Ticket coverage with (left side) and without (right side) irrelevant methods according to RQ2. The ratio of methods executed exclusively through tests increased. Colors: light gray = method is executed exclusively through test case; dark gray = method is covered during system startup; black = untested methods (i.e., open test gaps).}
	\label{Fig_uninterestingMethodsExcluded}
\end{figure}

In the conducted study, 28 of 110 test gaps (25.4\%) were marked as uninteresting.
To investigate the effect of blacklisting these methods, we recalculated the ticket coverage for the corresponding tickets.
The results can be seen in Figure \ref{Fig_uninterestingMethodsExcluded}.
As expected, the ticket coverage values increased after removing the noise.
As a consequence, less incompletely tested tickets remain which need to be investigated manually.

In mission-critical systems, excluding trivial methods might not be appropriate.

\medskip

\textbf{RQ3:}
We consider the coverage achieved during the system startup as acceptably low in the investigated study object.
Only 9.2\% of the covered methods were executed during the startup and in one half of the analyzed tickets the ticket coverage was completely test-dependent.

We found a high ratio of startup coverage in 5 tickets which were dealing with system parts that are related to the program initialization.
It was striking that the changesets of those tickets consisted of at most two methods.
For tickets that partially deal with initialization procedures, the influence of methods executed during the startup might be smaller.

If the possibility exists to activate the profiler after the initialization of the program (instead of activating it before the program start), it will not be necessary to determine the startup coverage.

The results of this research question show that some part of the achieved ticket coverage may be test-independent. The influence in our particular study, in which we executed and recorded each test case separately, was limited and did not distort the results.
However, it could be different in, e.g., exploratory testing or whenever coverage is not recorded in isolation for a ticket.

\subsection{Threats to Validity}

In this section, we discuss the threats to the internal, external, and construct validity of the study.

\medskip

\subsubsection{Internal Threats}
Threats to internal validity comprise reasons why the results could be invalid for the study object.

The responses gained from the developer survey regarding the relevance of the presented test gaps are a threat to validity.
Developers may not share the same view on the assessment whether a method is test relevant.
We mitigated this threat by involving 20 different developers and assigned at most 3 tickets to each developer.

A threat to the current implementation of the metric computation are syntactic code changes
 that do not change the semantic of a method.
For example, the rename of an identifier within a method implies that the method was changed for a certain ticket,
 even though its semantic did not change.
Consequently, this leads to a one-sided error because the computed set of changed methods may be larger than it actually is;
 nevertheless ticket coverage reveals at least the relevant gaps.
A future version of the implementation should include a refactoring detection to filter out semantically unchanged methods,
 increasing the ratio of relevant test gaps.

Another threat is caused by the development process that did not use feature branches at that time,
 such that the implementation of a ticket is not isolated from other development activities.
Therefore, changes to methods of different tickets conducted at the same time may overlap.
As a consequence, a method that was changed within the scope of ticket \#1
 and then moved to another class during the development of ticket \#2 before completing ticket \#1
 would no longer be recognized as changed within \#1.
However, the moved method would still be recognized within \#2 though, thus no method will travel under the radar.

Like for all coverage metrics, a threat regarding the metric itself is that it considers which methods got \textit{executed} during the test.
It does not take into account whether these methods were \textit{tested} with appropriate test assertions (in terms of comparison against expected values).
Therefore, coverage metrics should be employed with caution.
They are not necessarily a meaningful indicator for test effectiveness, especially not for system tests
\cite{niedermayr2016teststellme}.

\medskip

\subsubsection{External Threats}
Threats to external validity concern the generalization of the study results.

The study was conducted for the Java code of the study object Teamscale.
The results for Teamscale may not be applicable to other open- and closed-source projects
 and may hold only for Java code.
Therefore, further studies are necessary to determine whether the results are generalizable.

The identified criteria for methods that are potentially not worth testing in RQ1 and RQ2 comprise another threat regarding generalization.
While we and the developers involved in the survey assume that these methods are less likely to contain faults,
testing may still be useful because even simple methods can be faulty.
Therefore, it is necessary to empirically investigate the fault-proneness of simple methods (such as short getters, setters and delegation methods) to be able to make valid decisions regarding these methods.
Furthermore, the goal of testing and the impact of faults should be considered when reflecting about filtering out methods;
a company that is developing safety-critical software and aims to identify as many faults as possible
should take another decision than a company that wants to reduce the costs of testing.

\medskip

\subsubsection{Construct Validity}
Threats to construct validity concern the relationship between theory and observation, i.e., how accurate we measure the studied concept.

A threat to construct validity is that we defined the specifications for the manual test cases ourselves.
Therefore, the test cases may not have been proper or complete to test the associated ticket.
To mitigate this threat, we asked the ticket assignees to verify the developed test cases and adjusted the test cases according to their feedback.

\section{Conclusion and Future Work}
\label{Sec_Conclusion}

We presented the ticket coverage metric which expresses how well the changes conducted for a certain ticket are covered by tests.
The metric helps testers and their managers to get an overview over the testing state of tickets.
Furthermore, the metric can point to test gaps, i.e., untested changes in the source code, so that they can be tackled by testers.

The conducted empirical study confirmed that the revealed test gaps are relevant and useful for testers.
We identified that some of the test gaps were caused by incomplete test cases, a problem that occurs in many real world systems.
The study results also suggested how our first implementation of the metric can be improved:
We identified methods that are less likely to contain faults and may therefore not warrant being testing.
We classified them into groups and provided suggestions on how to exclude them automatically to gain a more expressive ticket-coverage result.
Finally, we presented how much coverage will be gained from startup routines if it is not possible to isolate the program startup from the coverage recording,
 and showed that its influence was negligible for our study object.

For future work, we intend to apply a refactoring detection.
It will allow us to exclude automatically methods that were not changed semantically from the ticket-coverage computation
 and thus, bring real changes into focus.
Furthermore, we want to replicate the study with further study objects, further programming languages,
 and the inclusion of coverage information from automated tests aside manual tests.
Finally, we plan to research whether ticket coverage is also a useful support in an exploratory testing process.

\section*{Acknowledgment}
This work was partially funded by the German Federal Ministry of Education and Research (BMBF), grant ``Q-Effekt, 01IS15003A''.
The responsibility for this article lies with the authors.

We thank Nils G\"ode for his valuable review.

\balance

\bibliographystyle{ieeetr}
\bibliography{bib/references} 

\end{document}